# The Gravitational Constant as a quantum mechanical expression

**Engel Roza**
Stripperwei 1, 5551 ST Valkenswaard, The Netherlands
Email: engel.roza@onsbrabantnet.nl

**Abstract**. A quantitatively verifiable expression for the Gravitational Constant is derived in terms of quantum mechanical quantities. This derivation appears to be possible by selecting a suitable physical process in which the transformation of the equation of motion into a quantum mechanical wave equation can be obtained by Einstein's geodesic approach. The selected process is the pi-meson, modeled as the one-body equivalent of a two-body quantum mechanical oscillator in which the vibrating mass is modeled as the result of the two energy fluxes from the quark and the antiquark. The quantum mechanical formula for the Gravitational Constant appears to show a quantitatively verifiable relationship with the Higgs boson as conceived in the Standard Model.

1.Introduction

The basic concept in quantum physics is the particle wave duality, which implies that a particle can dually be described by a mechanical equation of motion and by a quantum mechanical wave function. This wave function is the solution of a wave equation. The wave equation is obtained by a transformation of the particle's equation of motion in a way as conceived by Dirac [1]. Although Einstein's geodesic equation of motion [2] is the most generic of all, it is, so far, not adopted as the axiomatic base for the particle wave duality description. This is probably due to the mathematical complexity of 4D space-time. It is also the reason for the failure to unify quantum physics with gravity. Instead, Dirac derived his wave equation from the Einsteinean energy relationship of a particle in motion. Therefore, the equation is relativistic, but only special relativistic and not general relativistic. In this article I wish to develop the particle wave duality on the basis of Einstein's geodesic equation in 2D space-time and to compare it with Dirac's result. Next to that, I wish to motivate that the 2D space-time approximation is a justified modeling of a realistic physical process. This enables to relate gravity and quantum physics to the extent that the Gravitational Constant can be formulated as a quantitatively verifiable expression of quantum physical quantities. The concepts as will be outlined in this article, are invoked from previous work by the author [3]. Section 4 is devoted to a comparison of the 2D quantum mechanical wave equations as derived respectively by the geodesic approach (section 2) and Dirac's approach (section 3). This will result into an expression for the Gravitational Constant (section 5). In section 6 a physical process is selected that will deliver the quantity values. The result is subject to a relativistic correction (section 7). The final result is discussed in sections 8 and 9.

The equations in this article will be formulated in scientific notation and the quantities will be expressed in SI units. Space-time will be described on the basis of the "Hawking" metric (+,+,+,+).

## 2. The geodesic approach towards a 2D wave equation

Let us consider two frames of Cartesian coordinates $(x_1, x_2) \equiv (x, ict)$ and $(\xi_1, \xi_2) \equiv (\xi, ic\tau)$. The first one is the frame of a stationary ("lab frame") observer O. The second one is the ("center of mass") frame of an observer co-moving with a particle P. Time $t'$ and proper time $\tau'$ are normalized on the vacuum light velocity $c$ such that $t' = ict$ and $\tau' = ic\tau$, where $i = \sqrt{-1}$. Under stationary conditions, the 2D geodesic equation can be written as [4],

$$\frac{d^2 x}{d\tau^2} + \frac{1}{2g_{xx}} \left\{ \frac{\partial g_{xx}}{\partial x} \left(\frac{dx}{d\tau}\right)^2 - \frac{\partial g_{tt}}{\partial x} \left(\frac{dt'}{d\tau}\right)^2 \right\} = 0,$$

$$\frac{d^2 t'}{d\tau^2} + \frac{1}{g_{tt}} \frac{\partial g_{tt}}{\partial x} \left(\frac{dx}{d\tau} \frac{dt'}{d\tau}\right) = 0. \tag{1}$$

The quantities $g_{xx}$ and $g_{tt}$ are elements of the metric tensor. They determine the way how the frame $(\xi, \tau')$ of the co-moving observer – by considering functions $\xi(x, t')$ and $\tau(x, t')$ - is transformed into the frame $(x, t')$ of the stationary observer. In particular

$$\left(\frac{\partial \xi}{\partial x}\right)^2 + \left(\frac{\partial \tau'}{\partial x}\right)^2 = g_{xx} \quad \text{and} \quad \left(\frac{\partial \xi}{\partial t'}\right)^2 + \left(\frac{\partial \tau'}{\partial t'}\right)^2 = g_{tt}. \tag{2}$$

The adopted stationary condition means that the metric tensor components are supposed to be independent of time $t$. The geodesic equation expresses that the straight space-time path $(\xi, \tau)$ of the co-moving observer, is observed as a curved path by the lab frame observer. It is the consequence of expressing a presupposed field of forces into a metric tensor. The second part of (1) can be integrated into

$$\frac{dt'}{d\tau} = \frac{k_{int}}{g_{tt}}, \tag{3}$$

where $k_{int}$ is an integration constant. Considering that $dt/d\tau = 1$ in flat space-time ($g_{tt} = 1$), we have $k_{int} = ic$. In the case of a conservative field of forces, the local space-time interval is invariant, i.e., $d\tau'^2 = g_{xx} dx^2 + g_{tt} dt'^2$, so that

$$\left(\frac{d\tau'}{d\tau'}\right)^2 = g_{xx}\left(\frac{dx}{d\tau'}\right)^2 + g_{tt}\left(\frac{dt'}{d\tau'}\right)^2. \tag{4}$$

From (4) and (3) we get

$$\left(\frac{dx}{d\tau}\right)^2 = \frac{c^2}{g_{xx}}\left(\frac{1}{g_{tt}} - 1\right). \tag{5}$$

This result is consistent with the integration of the first part of (1) under consideration of (3).

Under the axiomatic quantum mechanical hypothesis

$$p_\mu \to \hat{p}_\mu \psi, \text{ with } \hat{p}_\mu = \frac{\hbar}{i}\frac{\partial}{\partial x_\mu}, \text{ where } p_\mu = m_0 \frac{dx_\mu}{d\tau}, \tag{6}$$

where $m_0$ is the rest mass of the particle in consideration, and where $\hbar$ is Planck's (modified) constant, we get from (3) and (5),

$$i\hbar \frac{\partial \psi}{\partial t} = \frac{m_0 c^2}{g_{tt}}\psi \quad \text{and} \quad \frac{\partial \psi}{\partial x} = \pm i m_0 f(x)\psi,$$

where $f(x) = \frac{1}{\hbar}\sqrt{\frac{c^2}{g_{xx}}(\frac{1}{g_{tt}}-1)}$. \hfill (7)

The two parts of (7) can be joined to a semantically correct quantum mechanical wave equation. This can be done by differentiation of the second part and addition to the first part after multiplying with $\hbar^2/2m_0$, resulting in,

$$i\hbar \frac{\partial \psi}{\partial t} + \frac{\hbar^2}{2m_0}\frac{\partial^2 \psi}{\partial x^2} = \{\frac{m_0 c^2}{g_{tt}} - (m_0 f^2 \pm i\frac{df}{dx})\}\psi. \tag{8}$$

The equation is temporally of first order and spatially of second order. It guarantees the positive definiteness of the wave function. This means that the spatial integral of squared absolute value of the wave function is time-independent. This is required to obey the semantics of the wave function, which states that its squared absolute value expresses the probability that the particle is at a certain moment at a certain place. As this certain place must be somewhere, the spatial integral has to be a time-independent. Eq. (8) has a similar format as Schrödinger's equation. The difference is in the right-hand part. In Schrödinger's equation this part is the (spatially dependent) potential energy of the particle in motion. Here, it is replaced by a quantity that expresses the metric curvature of space-time.

**3. Dirac's approach towards a 2D wave equation**

Dirac derived his relativistic quantum mechanical wave equation for a particle moving in free space from a heuristic elaboration of Einstein's energy relationship

$$E_W = \sqrt{(m_0 c^2)^2 + (c|\mathbf{p}|)^2}, \tag{9}$$

where $\mathbf{p}$ is the three-vector momentum ($ds/dt$, not to be confused with $p$), squared as

$$E_W^2 = -p_4^2 = m_0 c^2 + p_3^2, \tag{10}$$

or equivalently,

$$p_4'^2 + p_4'^2 + 1 = 0, \text{ with } p_\mu' = \frac{p_\mu}{m_0 c}. \tag{11}$$

Under consideration of the required semantics, Dirac expanded and transformed this quadratic equation into a set of two linear ones,

$$[\sigma_1]\begin{bmatrix}\hat{p}'_4\psi_0\\\hat{p}'_4\psi_1\end{bmatrix}+[\sigma_3]\begin{bmatrix}\hat{p}'_3\psi_0\\\hat{p}'_3\psi_1\end{bmatrix}+[\sigma_2]\begin{bmatrix}\psi_0\\\psi_1\end{bmatrix}=0,$$

where $[\sigma_i]$ are the Pauli matrices, so that

$$\hat{p}'_4\psi_1+\hat{p}'_3\psi_0-i\psi_1=0 \quad \text{and} \quad \hat{p}'_4\psi_0-\hat{p}'_3\psi_1+i\psi_0=0. \tag{12}$$

If the velocity of the particle is small with respect to the light velocity, we have $m_0c^2 \gg p_3^2$, so that from (10),

$$p_4 \approx \pm i m_0 c, \text{ and therefore } \hat{p}'_4\psi_1 \approx \pm i\psi_1. \tag{13}$$

From (13) and the first part of (12), and adopting the minus sign to avoid a meaningless result, it follows

$$\psi_1 \approx \frac{i}{2}\hat{p}'_3\psi_0. \tag{14}$$

After substitution of (14) into the second part of (12), we get, under consideration of (6),

$$i\hbar\frac{\partial\psi_0}{\partial t}+\frac{\hbar^2}{2m_0}\frac{\partial^2\psi_0}{\partial x^2}-m_0c^2\psi_0=0. \tag{15}$$

Mutatis mutandis, and now adopting the plus sign in (13), we get

$$i\hbar\frac{\partial\psi_1}{\partial t}+\frac{\hbar^2}{2m_0}\frac{\partial^2\psi_1}{\partial x^2}-m_0c^2\psi_1=0 \quad \text{and} \quad \psi_0 \approx -\frac{i}{2}\hat{p}'_3\psi_1. \tag{16}$$

Eqs (15) and (16) can be summarized as

$$i\hbar\frac{\partial\psi_1}{\partial t}+\frac{\hbar^2}{2m_0}\frac{\partial^2\psi_1}{\partial x^2}-m_0c^2\psi_1=0 \quad \text{and} \quad \psi_0 \approx \pm\frac{\hbar c}{m_0c^2}\frac{\partial\psi_1}{\partial x}. \tag{17}$$

Eq. (17) is known as the Pauli-Schrödinger approximation of Dirac's equation. The solution is a two-component wave function, with a dominant component and a minor (spin) component that can assume two states.

If a particle is subject to a field of forces, Dirac's Equation as formulated in (17) does not hold. Potentially however, its format can be preserved if the operators on the wave function are redefined in a suitable way. This is known as the application of the Principle of Covariance. This involves a redefinition of the operators in the wave equation, implying that derivatives of the wave function are replaced by covariant derivatives. In particular,

$$\partial_\mu\psi \to D_\mu\psi = (\partial_\mu+\frac{gA_\mu}{\hbar})\psi. \tag{18}$$

where $g$ is a dimensionless generic coupling factor and where, generically, $A_\mu$ are the components of the four-vector potential $(A_1, A_2, A_3, A_4)$ that characterizes the field forces, and where $A_4 = i\Phi/c$ is the scalar part of the field. In the case that the field is characterized by a scalar only, we get for the dominant wave component from (17) and (18),

$$i\hbar \frac{\partial \psi}{\partial t} - g\Phi\psi + \frac{\hbar^2}{2m_0} \frac{\partial^2 \psi}{\partial x^2} - m_0 c^2 \psi = 0. \tag{19}$$

**4. Comparing the geodesic approach with Dirac's approach**

Comparing (19) with (8), it is concluded that, under non-relativistic conditions, the particle's wave equation in a conservative scalar field of forces derived by the geodesic approach and Dirac's approach, are equivalent if

$$g\Phi + m_0 c^2 = \frac{m_0 c^2}{g_{tt}} - (m_0 f^2 \pm i \frac{df}{dx}), \tag{20}$$

where $f$ is a function of the metric components $g_{tt}$ and $g_{xx}$ (not to be confused with the quantum mechanical coupling factor $g$).

Let us proceed by considering conditions as often assumed in general relativistic gravitational problems. These are, *isotropy* and *weak field* condition, implying

$$g_{tt} = 1 + \Delta(x), \ g_{xx} = 1 + \Delta_2(x), \text{ where } |\Delta x|, |\Delta x_2| << 1. \tag{21}$$

This enables to rewrite (20) as

$$g\Phi = \pm \frac{\hbar c}{4\sqrt{\Delta}} \frac{\partial \Delta}{\partial x} + \frac{m_0 c^2}{2} \Delta. \tag{22}$$

This expression relates the space-time curvature with the potential energy of the particle in motion. In the particular case that the potential field shows a spatial quadratic dependency, the wave equation represents a quantum mechanical oscillator. This is true if

$$g\Phi = g\Phi_0 (k_0 + k_2 \lambda^2 x^2), \tag{23}$$

where the normalization parameter $\lambda$ is introduced to make the constant $k_2$ dimensionless. Elementary algebra shows that (20) and (22) are equivalent if

$$\Delta = \beta^2 x^2, \text{ where } \beta = \frac{2k_0 g\Phi_0}{\hbar c}, \text{ provided that}$$

$$g\Phi_0 = \frac{k_2 \lambda^2}{k_0^2} \frac{(\hbar c)^2}{m_0 c^2}. \tag{24}$$

Eq. (24) imposes an interrelationship between the particle's potential energy and its rest mass. This may seem an over-constraint. Actually, it is not, because the rest mass energy curves space-time and the potential energy is a manifestation of the space-time curvature (otherwise equating the geodesic approach with Dirac's approach would be meaningless).

**5. The Gravitational Constant**

Energetic fields are formats of energy, similarly as massive particles in rest. All formats of energy are subject to Einstein's field equation. Accepting the universality of this principle beyond the realm of gravity, justifies applying Einstein's field equation to nuclear energetic fields. One side of this equation is the Einstein tensor ($G_{\mu\nu}$), which is expressed in quantities that can be derived from the metric tensor. The other side is the energy momentum tensor ($T_{\mu o}$), which contains quantities that can be derived from the energy present in the space as characterized by the metric tensor. There is a proportionality factor $G$ involved, known as the gravity constant. The full expression is

$$G_{\mu\nu} = \frac{8\pi G}{c^4} T_{\mu\nu} \quad \text{with} \quad G_{\mu\nu} = R_{\mu\nu} - \frac{1}{2} R g_{\mu\nu}. \tag{25}$$

Here, $R_{\mu\nu}$ and $R$ are respectively the so-called Ricci tensor and the Ricci scalar, which can be calculated if the metric tensor components $g_{\mu\nu}$ are known [1,4]. Let us start by calculating the energy momentum tensor. This presupposes knowledge of the spatial energy density. We wish to proceed under the assumption that all energy is comprised in the potential energy $g\Phi$ of the moving object and that the energy density $w$ is given as (see Appendix I),

$$w = \frac{1}{8\pi(\hbar c)} \left|\frac{d\Phi}{dx}\right|^2. \tag{26}$$

The viability of this assumption will be shown for a particular physical process, to be discussed in section 6. From (26), it follows that

$$T_{11} = T_{22} = \frac{1}{8\pi(\hbar c)} \left|\frac{d\Phi}{dx}\right|^2 \quad \text{and} \quad T_{12} = T_{21} = 0. \tag{27}$$

Note that the index $\mu = 1$ applies to the spatial dimension and $\mu = 2$ to the temporal dimension. It follows straightforwardly from (27) and (23) that,

$$T_{11} = \frac{\Phi_0^2}{8\pi(\hbar c)} (2k_2 \lambda^2 x)^2. \tag{28}$$

The more difficult part is the calculation of the Einstein tensor from the Ricci tensor $R_{11} (= R_{22})$ and the Ricci scalar $R$, to be obtained from the metric tensor $g_{11} (= g_{22})$. Because of the 2D isotropy condition, the calculation results into rather simple expressions [4],

$$R_{11} = R_{22} = -\frac{1}{2} \frac{g_{11}''}{g_{11}} + \frac{1}{2} \frac{g_{11}'^2}{g_{11}^2}, \text{ and} \tag{29}$$

$$R = \sum_{\mu=1}^{2} \sum_{\nu=1}^{2} g^{\mu\nu} R_{\mu\nu} = 2 g_{11} R_{11}. \tag{30}$$

Note: $a'$ and $a''$ is short for differentiation and double differentiation (after $x$ )of the parameter $a$, and $g^{\mu\nu}$ is the inverse of the matrix $g_{\mu\nu}$. Application of (21) and (23) on (29) and (30), gives

$$R_{11} = -\beta^2 + 2\beta^4 x^2, \tag{31}$$

so that

$$G_{11} = G_{22} = R_{11} - \frac{1}{2}Rg_{11} = R_{11}(1 - g_{11}^2) = R_{11}\{1 - (1 + 2\Delta)\} = -2\Delta R_{11} = 2\beta^4 x^2. \tag{32}$$

From (25),(28) and (32), under consideration of (24), the Gravitational Constant $G$ follows as,

$$G = \frac{(\hbar c)c^4}{2\Phi_0^2 k_2^2}(\frac{\beta}{\lambda})^4 \quad (\frac{[m^3]}{[kg][s^2]}), \text{ with } \beta = \frac{2k_0 g\Phi_0}{\hbar c}. \tag{33}$$

This results means that it is possible to express the Gravitational constant into quantum mechanical quantities, provided that we can identify a physical process that can be modeled as a harmonic 2D quantum mechanical oscillator where the mass of the object in motion is extracted from the field's potential energy.

**6. The physical process**

In this section, it will be claimed that such a process exists. It is the basic quark dipole, known as the pi-meson. It is commonly accepted that the quark masses have their origin from an omni-present scalar field of energy, known as the Higgs field. This field is functionally described by a Lagrangian density $\mathcal{L}$ with two characteristic quantities $\mu_c$ and $\lambda_c$, such that [7, p.363]

$$\mathcal{L} = -\frac{1}{2}(\nabla\Phi)^2 + U_H(\Phi) + \rho\Phi, \text{ with } U_H(\Phi) = -\frac{1}{2}\mu_c^2\Phi^2 + \frac{1}{4}\lambda_c^2\Phi^4. \tag{34}$$

Usually, the source term $\rho\Phi$ is omitted, because it is simply stated that an unknown source sustains the field. But let us consider the consequences if we do not wish to accept an incomplete Lagrangian density description. To do so, let us compare the Lagrangian density of the Higgs field with the Lagrangian density of the type

$$\mathcal{L} = -\frac{1}{2}(\nabla\Phi)^2 + U(\Phi) + \rho\Phi, \text{ with } U(\Phi) = \frac{\lambda^2}{2}\Phi^2. \tag{35}$$

Let us further suppose that the source is a three-dimensional Dirac distribution $4\pi\Phi_0\delta(r)$. Application of the Lagrange-Euler equation yields a differential equation for the spatial behavior of the field's potential energy . In this case,

$$\frac{1}{r}\frac{d^2}{dr^2}(r\Phi) + \lambda^2\Phi = -4\pi\Phi_0\delta(r), \tag{36}$$

The solution of (36) is

$$\Phi = \Phi_0 \frac{\exp(-\lambda r)}{\lambda r}. \tag{37}$$

If $\lambda \to 0$ and $\Phi_0 = Q\lambda/(4\pi\varepsilon_0)$, where $Q$ is the electric charge of a pointlike source and $\varepsilon_0$ is the free space electric permeability, this expression represents a Coulomb field. Generically it represents a field with a format that corresponds with the potential as proposed by Yukawa [5] to explain the short range of a nuclear force. It can also be viewed as a screened Coulomb field, which originates if the free Coulomb flux is suppressed by a surrounding space charge, such as first described by Debije [6]. The Debije shielding is observed in a wide variety of physical processes, particularly in plasma physics and in astrophysics. In these processes the Debije shielding length $1/\lambda$ may range from values as large as $10^5$ m to values as small as $10^{-9}$ m.

Unfortunately, the high non-linearity of the potential energy term $U_H(\Phi)$ in the Lagrangian density of the Higgs field, as shown, in (34) prevents the analytical derivation of a spatial description of the Higgs field from a pointlike source $\rho = 4\pi\Phi_0\delta(r)$. But if we cannot solve the spatial field equation analytically, why not solving it numerically? The way how to do has been documented in previous work [3a], in which it has been shown that the potential $\Phi(r)$ that satisfies the wave equation derived from (34), is closely approximated by,

$$\Phi(r) = \Phi_0 \frac{\exp[-\lambda r]}{\lambda r}\left(\frac{\exp[-\lambda r]}{\lambda r}-1\right) \text{ with } \frac{1}{2}\mu_c^2 = 1.06\lambda^2 \text{ and } \frac{1}{4}\lambda_c^2 = 32.3\frac{\lambda^2}{\Phi_0^2}. \tag{38}$$

This field shows *far field* characteristics corresponding to the Yukawa ones, next to *near field* characteristics with opposite direction that represents a short-range force. We identify the far field force as the force that is commonly known as the *weak force* (short for force responsible for *weak interaction*) and the scalar near field force as the *strong force* (short for force responsible for *strong interaction*). The *nuclear force* is the combination of the two, such that the field, as shown by (38), is given by

$$\Phi = \Phi_s - \Phi_w \text{ with } \Phi_s = \Phi_0 \frac{\exp[-2\lambda r]}{(\lambda r)^2} \text{ and } \Phi_w = \Phi_0 \frac{\exp[-\lambda r]}{\lambda r}. \tag{39}$$

If we identify the quark as the source of this nuclear field, a nice picture is obtained. Any quark couples to the field of any other quark with the coupling factor $g$, such that it feels a nuclear force $F$ described as

$$F = -g\frac{\partial \Phi}{\partial r}, \tag{40}$$

where the quantum mechanical coupling is supposed to be equal to the square root of the electromagnetic fine structure constant ($g^2 \approx 1/137$, see Appendix I). This means that any quark is repelled by any other quark under influence of the far field, but attracted by the near field. As a consequence, structures are possible to exist that are composed by two quarks or three quarks, which are holding each other in a stable equilibrium. Similarly as Lorentz did in the past for electrons, we may suppose that the masses of these structures will primarily, if not all, be determined by the fields that result from the potential fluxes from the composing quarks. In this model, two quarks, positioned at a spacing $2d$ apart, compose a structure, the center of mass of which will vibrate around the position just half-way the two quarks. Therefore, this quark dipole, to be identified as meson, can be modeled effectively as an equivalent single-body quantum mechanical oscillator with a certain effective mass that is built by the energetic fluxes from the quarks.

The far field component (weak force) can be conceived as the scalar part of the four-vector potential $(A_1, A_2, A_3, A_4)$, of a field with the characteristics of Proca's generalization of a Maxwell field, described by a Lagrangian density of the type,

$$\mathcal{L} = -\frac{1}{4} F^{\mu\nu} F_{\mu\nu} + \frac{1}{2} \lambda^2 A^\nu A_\nu + J_\nu A^\nu \quad \text{with} \quad F_{\mu\nu} = \frac{\partial A_\mu}{\partial x_\nu} - \frac{\partial A_\nu}{\partial x_\mu}, \tag{41}$$

where the vector components $J_\mu$ represent the sources of the field, possibly exclusively consisting of the pointlike well $\rho = 4\pi\Phi_0 \delta(r)$. In the past, such a field has been considered as a candidate to explain the origin of nuclear forces indeed. It has been abandoned because the $\lambda^2 -$ term spoils the gauge constraint that is required to obtain the covariant format of Dirac's equation. Stueckelberg, however, has shown that the gauge constraint remains valid in the case that the vectorial Proca field is supplemented by an additional scalar field. In Stueckelberg's time, there was no rationale for this artificial escape. Within the context of this article it makes sense, because the scalar near field in (39) may serve the purpose. The justification for it is shown in Appendix II. A side effect of this solution is the removal of a possible renormalization problem associated with the inverse square potential format of the scalar near field.

The meson is subject to a quantum mechanical wave equation, which will be developed within the center of mass frame. As noted before, the two constituting quarks will hold each other in a stable equilibrium. This enables us to develop a one-body equivalent of a two-body oscillator. Although such oscillator resembles a classical one, there is a fundamental difference. In the classical case, we have two masses and a (potential) field in between. In the classical case, the energy captured in the two masses is much larger than the energy represented by the field. In the model to be developed here, the other extreme is adopted: the bare mass of the bodies is supposed to be negligible as compared with the field energy. It makes the oscillator relativistic, because the mass in the wave equation is no longer the mass of the two bodies, but it is an equivalent mass that captures the energy of the field. In spite of the relativistic nature of the model, the center of mass view allows applying the Pauli-Schrödinger approximation of Dirac's wave equation, describing a linear motion of an effective mass between the quark centers. Therefore, we write,

$$-\frac{\hbar^2}{2m_m} \frac{d^2\psi}{dx^2} + \{U(d+x) + U(d-x)\}\psi = E\psi. \tag{42}$$

Here, $m_m$ the non-relativistic effective mass of the center, $V(x) = U(d+x) + U(d-x)$ its potential energy and $E$ the generic energy constant, which will be subject to quantization. From (32) we have,

$$U(x) = g\Phi_0 \left( \frac{\exp(-2\lambda x)}{(\lambda x)^2} - \frac{\exp(-\lambda x)}{\lambda x} \right). \tag{43}$$

The potential energy $V(x)$ can be expanded as

$$V(x) = U(d+x) + U(d-x) = g\Phi_0 (k_0 + k_2 \lambda^2 x^2 + ....), \text{ where} \tag{44}$$

$$k_0 = 2\left( \frac{\exp(-2\lambda d)}{(\lambda d)^2} - \frac{\exp(-\lambda d)}{\lambda d} \right), \text{ and}$$

$$k_2 = \frac{\exp(-2\lambda d)}{(\lambda d)^5}(6 + 4\lambda^2 d^2 + 8\lambda d) - \frac{\exp(-\lambda d)}{(\lambda d)^2}(2 + \lambda d + \frac{2}{\lambda d}). \tag{45}$$

The two quarks in the meson settle in a state of minimum energy, at a spacing $2\lambda d = 2d'_{min}$, determined from the condition

$$\frac{\exp(-\lambda d)}{\lambda d} = \frac{1}{2}, \tag{46}$$

so that $d'_{min} = \lambda d = 0.8561$; $k_0 = -1/2$ and $k_2 = 2.36$ (at $\lambda d = d'_{min}$).

The quantum mechanical oscillator as represented by (42) is subject to excitation. This means that the energy constant $E$ is subject to excitation. Under the approximation of the potential energy by a quadratic polynomial as (44), the constants associated with the vibration energy $E_n = (n + 1/2)\hbar\omega$, are equally spaced. The frequency $\omega$ is given by the generic basic relationship [9],

$$\frac{m_m \omega^2}{2} = g\Phi_0 k_2 \lambda^2. \tag{47}$$

A particle-antiparticle conjunction, as in the case of a meson, has particular characteristics. All mass is in the binding energy between the two particles. This allows the conclusion that the spacing $2d$ between the quark and the antiquark is determined by half the wavelength of a single harmonic energetic standing wave (boson) with *phase velocity c*, so that

$$\lambda = \frac{2(\hbar\omega_W)d'_{min}}{\alpha\pi(\hbar c)}, \tag{48}$$

where $\alpha$ is a dimensionless constant with order of magnitude 1, introduced for corrections because of the crude modeling and where $\hbar\omega_W$ is the energy of a bosonic mass particle responsible for change of energetic states of the meson. This boson is known as the weak-interaction boson. Its energetic value is known in the lab frame from experimental evidence of nuclear decay processes, but will be subject to relativistic correction in the center of mass frame. From (47) (24) and (48) it follows that

$$\hbar\omega_W = 2|k_0|g\Phi_0. \tag{49}$$

(Note: the value of the effective mass $m_m$ is irrelevant within the scope of this paper. Eventually it comes manifest in the lab frame as the rest mass of the meson).

Under consideration of (48) and (49), the gravitational constant, as expressed by (33), can thus be written as,

$$G = \frac{2g^2 k_0^2 (\hbar c)c^4}{(\hbar\omega_W)^2 k_2^2} \left(\frac{\alpha\pi}{2d'_{min}}\right)^4 \quad \left(\frac{[m^3]}{[kg][s^2]}\right). \tag{50}$$

This analysis has been made in the center of mass frame, i.e., the frame of the co-moving observer. The energy $\hbar\omega_W$ is known in the lab frame as $m'_W = 80.4$ GeV. The co-moving observer experiences the lab frame velocity as the lab frame speed of the pi-meson. Owing to this particle-antiparticle structure, the pi-meson flies at near light velocity $v_\pi \approx c$. Therefore, (50) is subject to a major relativistic correction, so that, effectively,

$$G = \frac{2g^2 k_0^2 (\hbar c) c^4}{m_W^{'2} k_2^2} \left(\frac{\alpha\pi}{2d'_{min}}\right)^4 \Delta_\pi \quad \left(\frac{[m^3]}{[kg][s^2]}\right), \text{ with } \Delta_\pi = 1 - (v_\pi/c)^2. \tag{51}$$

The relativistic correction will be considered in the next section.

**7. The relativistic correction**

The composite field of three quarks in a baryon, relatively far from the internal structure, as built-up by three contributions of the type as defined by (39), shows the same behavior as that of a single quark. It has both the characteristics of the attracting near field and those of the repulsive far field. If the *inter-baryon* interaction is represented by a massive bosonic particle with a certain spatial range and if this range is one or more orders of magnitude larger than the range of *the intra-baryon* interaction bosons, the inter-baryon behavior is just a scaled behavior of the intra-baryon behavior. In that case a similar potential function can be defined for the remnant field of the baryon similarly as (39), with the only difference that the $\lambda$-value has to be scaled. Therefore, supposing that $\lambda \gg \lambda_\pi$, the inter-baryon field can be defined as

$$\Phi_b(r) = \Phi_0 \frac{\exp(-\lambda_\pi r)}{\lambda_\pi r}\left(1 - \frac{\exp(-\lambda_\pi r)}{\lambda_\pi r}\right). \tag{52}$$

Other baryons may couple to this field, with some dimensionless coupling factor $g'$. Identifying the interacting bosons as pions with rest mass energy $\hbar\omega_\pi = m'_{0\pi}$ (= 135.0/139.6 MeV), we have

$$\lambda_\pi = \frac{2m'_{0\pi} d'_{min}}{\alpha\pi(\hbar c)}. \tag{53}$$

According to this model, in non-excited state, the distance $2d$ between bound baryons therefore is, typically,

$$2d \approx 2\lambda_\pi d'_{min} \ (\approx 1.16 \text{ fm}). \tag{54}$$

The pions travel at near light speed. As shown by Watkins [10], this speed $v_\pi$ can be determined from the half life value $t_0$, in proper time denoted as $\tau_0$. It is based upon the relationship between the temporal decay rate $\gamma$ and the spatial decay parameter $\lambda_\pi$. The decay behavior of $N(\tau)$ particles in proper time is given by

$$N(\tau) = N_0 \exp(-\gamma\tau). \tag{55}$$

From the half life definition, given by

$$N(\tau + \tau_0) = N(\tau)/2 \tag{56}$$

It follows from (55) and (56) that

$$\tau_0 = \frac{\ln(2)}{\gamma}. \tag{57}$$

The relationship between spatial decay and temporal decay is frame independent and given by

$$\gamma = \lambda_\pi v_\pi \tag{58}$$

so that from Eqs. (55)- (58) after relativistic correction for $\tau_0$,

$$t_0 = \frac{\ln(2)}{\lambda_\pi v_\pi \sqrt{\Delta_\pi}}, \text{ where } \Delta_\pi = 1 - (v_\pi/c)^2, \tag{59}$$

so that, for $v_\pi \approx c$, under consideration of (53),

$$\sqrt{\Delta_\pi} = \sqrt{1-(v_\pi/c)^2} = \frac{\alpha\pi(\hbar c)\ln(2)}{2 d'_{min} m'_{0\pi} c t_0}. \tag{60}$$

**8. Result**

Summarizing, we have (51) and (60),

$$G = \frac{2 g^2 k_0^2 (\hbar c) c^4}{m'^2_W k_2^2} \left(\frac{\alpha\pi}{2 d'_{min}}\right)^4 \Delta_\pi \quad \left(\frac{[m^3]}{[kg][s^2]}\right), \text{ where}$$

$$\sqrt{\Delta_\pi} = \frac{\alpha\pi(\hbar c)\ln(2)}{2 d'_{min} m'_{0\pi} c t_0}.$$

The result of this analysis is summarized in table I. The left-hand column shows the values of the physically known quantities. The middle column shows the known dimensionless constants as they are established in this theory. The right-hand column shows that the known value of the gravitational constant is obtained for the parameter value $\alpha = 0.69$. This value is in agreement with the presupposed order of magnitude 1.

| physics | this theory | calculated |
|---|---|---|
| $\hbar c = 193$ MeV fm | | $\alpha = 0.69$ |
| $m'_{0\pi} = 139.6$ MeV | $k_2 = 2.36$ | $\sqrt{\Delta_\pi} = 1.59 \times 10^{-16}$ |
| $t_{0\pi} = 2.603 \times 10^{-8}$ s | $k_0 = -1/2$ | $G = 6.67 \times 10^{-11}$ m$^3$kg$^{-1}$s$^{-2}$ |
| $m'_W = 80.4$ GeV | $d'_{min} = 0.8526$ | |
| $g^2 = 1/137$ | | |

**Table I.** Numerical result for the gravitational constant, calculated from quantum mechanical quantities.

In view of the order of magnitude of the quantities involved, this attempt to unify gravity with quantum mechanics yields a surprising fit of the theoretically calculated value of the gravitational constant with the experimentally established constant of nature.

The credibility of this result can be underlined with an interesting observation. It has to do with the 126,5 GeV particle, discovered in 2012 by CERN and identified as the "Higgs boson" of the Standard Model. The energetic value of its mass is known to be a function of the quantity $\mu_c$ in the functional definition of the Higgs field as given by (38). In particular [7, p.364]

$$m'_H = \mu_c(\hbar c)\sqrt{2}. \tag{61}$$

The Standard Model, however, is unable to establish theoretically based values for the two Higgs quantities $\mu_c$ and $\lambda_c$. That prevents a calculation of (61). Interestingly, though, the theory as outlined in this article, may do. How? Similarly as the boson of the electromagnetic field, neither the boson of vectorial far field, nor the boson of the scalar near field are observables. They can only show up by "signatures". There is no reason why the view on the Higgs field as developed in this article would exclude the possibility that photons interact with the Higgs field as spread by quarks. Such interactions would produce new particles of the same kind as predicted in the state-of-art theory. Let us try to calculate the mass equivalent of such new particles under application of the relationships as developed in this article.

As shown by (38), the equivalents of the quantities $\mu_c$ and $\lambda_c$ in the common functional description of the Higgs field are the spatial field quantities $\Phi_0$ (for is strength) and $\lambda$ (for its spatial range). It has been shown in Eq. (49) that the strength quantity can be determined from the weak interaction boson $\hbar\omega_W = m'_W$ (= 80.4 GeV), as

$$\Phi_0 = \frac{m'_W}{2g|k_0|}, \tag{62}$$

and that the spatial range quantity $\lambda$ follows from the ratio $\Phi_0/\lambda$, calculated from (48) and (49) as,

$$\frac{\Phi_0}{\lambda} = \frac{\alpha\pi(\hbar c)}{4g|k_{0m}|d'_{min}}. \tag{63}$$

(Note that, unlike the individual quantities $\Phi_0$ and $\lambda$, the numerical value of this ratio is frame-independent). It has to be emphasized that the results (62) and (63) have to be credited to the general relativistic view on the physical quark dipole. Applying (38) on (61), we have

$$m'_H \approx 2.\lambda(\hbar c). \tag{64}$$

From (62-64), and table I, it follows that,

$$m'_H = \frac{4d'_{min}m'_W}{\alpha\pi} = 127 \text{ GeV}. \tag{65}$$

This theoretically established energy value $m'_H$ of the "Higgs mass" is close to the measured value by CERN. Within the context of this article, it may be regarded as a confirmation of the viability of the relationship between gravity and quantum physics as expressed by (51), (60) and the numerical result given in Table I.

## 9. Discussion

It will be clear that the validation of a formula that expresses the Gravitational Constant in quantum mechanical quantities is an important contribution to the on-going challenge in present theoretical physics to relate gravity with quantum physics. It will also be clear that a claim that this can be done without conceiving a new theoretical framework beyond the established ones, will be regarded as controversial. Nevertheless, in this article an attempt to do so has resulted in a formula that can be easily numerically verified in an outcome that nicely fits with the well-known established experimental evidence. This numerical result cannot be denied. The theoretical steps might be subject to criticism, but the criticism should then answer the question why such a result is possible if something is wrong with the basic approach. So, let me summarize the approach, thereby stipulating different angles of view as compared to present theory. Among the various steps taken, there are three major ones. One of these is the view that Einstein's field equation is universal, therefore also valid beyond the realm of gravity and, therefore, applicable to nuclear energetic fields as well. The second one is the view that Dirac's approach for deriving a quantum mechanical wave equation from an equation of motion can be made more fundamental by taking Einstein's geodesic equation as a starting point rather than restricting it to Einstein's expression for relativistic energy. The third one has to do with the Higgs field. Similarly as in present-state quantum theory, the Higgs field is considered as a field from which energy can be subtracted for the purpose to give mass to particles. The view taken, however, is that the lack of a source term in the formulation of its Lagrangian density is unacceptable. Accepting a pointlike source as in conventional field theory, enables the derivation of a spatial field description, albeit that a numerical approach is required to do so. The logical step taken is, to identify the quark as the pointlike source of energy. This is not all. The most essential element in this third step is the split of the Higgs field into two different components: a bosonic vectorial far field in terms of Proca's generalization of the Maxwell field and an additional bosonic scalar near field with a narrow spatial reach. Owing to these characteristics, there is no need to explain the origin of massive nuclear bosons as a consequence of the interaction of mass less particles with a scalar field. Instead, the nuclear force bosons come forward by definition, similarly as photons show up in a Maxwell field.

Finally, I want to emphasize the difference between the $\lambda-$ term, which specifies the spatial range of the nuclear force, and a physical mass term $\lambda = mc^2/hc$, such as in Proca's original formulation. Owing to the invariance of the ratio $\Phi_0 / \lambda$, the energy of a "H-type" Proca boson remains, similar as a photon, the same in any inertial frame. If somebody tries to bring the Proca boson to rest, the $\lambda-$ term changes in coherence with the change of $\Phi_0$. It is for that reason that the quantum of the Higgs field cannot be identified as an observable massive boson. Experimental data on observables (like many fermions) are "hard", but experimental data on non-observables, like bosons flying at (near) light speed, are "soft". They show up as "signatures", which are interpreted with a theory in mind. A signature that supports the theory as developed in this article is the theoretically derived value for the mass of the 126,5 GeV "Higgs particle", which in present-state theory needs to be empirically established. This is a second major result of the analysis presented.

**Appendix I**

In the physical process as described in section 6, the nuclear field is conceived as an energetic field similar to that of an electromagnetic field. The spread of an energetic flux by pointlike wells causes creates a spatial field, which, in Maxwell's theory , is characterized by its energy density, i.e. the amount of field energy per unit of volume. The nuclear equivalent of the Maxwellian field density can be found on the basis of the unification hypothesis. Within the context of our physical process this hypothesis is, in SI units, formulated as:

$$e\nabla\Phi_e = g\nabla\Phi \quad \text{and} \quad e^2 = 4\pi\varepsilon_0 \hbar c g^2, \tag{A1}$$

where $\Phi_e$ and $\Phi$ are the scalar parts of respectively the electromagnetic potential and the nuclear potential, $e$ is the elementary electric charge and $\varepsilon_0$ the free space electric permeability. The hypothesis states that the square of the nuclear coupling factor $g$ is equal to the electromagnetic fine structure constant ($g^2 \approx 1/137$). The justification of the unification hypothesis has to be provided by experimental evidence. For its validation, see [3]. The energy density of the nuclear field can now expressed in similar terms as the electric energy density of the electromagnetic field, i.e., as

$$w = \frac{1}{2}\varepsilon_0 |\nabla\Phi_e|^2 = \frac{1}{8\pi\hbar c}|\nabla\Phi|^2. \tag{A2}$$

**Appendix II**

In 1938, Ernst Stueckelberg [11,12] showed that, under particular circumstances, the elegancy of the Principle of Covariance on the basis of minimum substitution, can be maintained for Proca type fields. This will be the case if, next to a Proca field, an auxiliary scalar bosonic field $B$, will be present, such that Proca's Lagrangian is modified into, [12],

$$\mathcal{L} = -\frac{1}{4}F^{\mu\nu}F_{\mu\nu} + \frac{1}{2}\lambda^2(A^\mu - \partial_\mu \frac{B}{\lambda})^2 - \frac{1}{2}(\partial_\mu A^\mu + \lambda B)^2. \tag{B1}$$

This modified Lagrangian density remains unaffected under the gauges,

$$A'_\mu = A_\mu - \frac{i\hbar}{g}\partial_\mu \vartheta_s(x_1, x_2, x_3, x_4)$$
$$B' = B + \frac{i\hbar}{g}\lambda \vartheta_s(x_1, x_2, x_3, x_4). \tag{B2}$$

Therefore, it is allowed to use the covariant Dirac equation in a Proca field of forces if an auxiliary scalar field is present as well. Let us try to identify the near field (strong force) as defined in (39) as Stueckelberg's auxiliary phantom scalar field. First of all, we have to cope with the $r^2$-term in the denominator of the near field expression, which is not compatible with a pointlike source. This would suggest that the source of the scalar field is a dipole rather than a monopole. Unfortunately, although the dipole shows an inverse square behavior of the energetic flux indeed, it shows an angular dependency as well. The escape comes from a reconsideration of the axiomatic principle as adopted by the author in his sequence of papers. It has to do with the numerical fit of the spatial expression (38) with the functional expression of the Higgs field. Curiously, another expression gives a fit with a similar accuracy. The fit is obtained by

$$\Phi(r) = \Phi_0\{a\frac{\exp(-p\lambda r)}{\lambda r} - b\frac{\exp(-q\lambda r)}{\lambda r}\}$$

with $a = 5.118$; $b = 0.737$; $p = 4.295$ and $q = 0.884$. (B3)

This can be rewritten as,

$$\Phi(r) = \Phi'_0 \{a' \frac{\exp(-p'\lambda' r)}{\lambda' r} - \frac{\exp(-\lambda' r)}{\lambda' r}\}, \tag{B4}$$

where $\lambda' = q\lambda$; $\Phi'_0 = \Phi_0 bq$; $p' = p/q$ and $a' = a/b$.

This allows to incorporate the sources of the vectorial far field and the scalar near field into a complete Lagrangian description. Moreover, we may invoke the generalization proposed by 't Hooft and Veltman, by including an additional real parameter $\gamma_0$, [12,13 p.76], such that,

$$\mathcal{L} = -\frac{1}{4}F^{\mu\nu}F_{\mu\nu} + \frac{1}{2}\lambda'^2(A^\mu - \partial_\mu \frac{B}{\lambda'})^2 - \frac{1}{2\gamma_0}(\partial_\mu A^\mu + \gamma_0 \lambda' B)^2 + J_\mu A^\mu + \rho B. \tag{B5}$$

Applying the Euler-Lagrange Equation yields two wave equations. Describing the pointlike sources for the far field and near field, respectively, as $\rho_W = 4\pi\Phi'_0 \delta(r)$ and $\rho_s = 4\pi a' \Phi'_0 \delta(r)$, and defining $B \equiv \Phi_s$, we get for the scalar part of the time independent (vector type) far field,

$$\frac{1}{r}\frac{d^2}{dr^2}(r\Phi_W) + \lambda'^2 \Phi_W = -4\pi\Phi'_0 \delta(r), \tag{B6}$$

and for the time independent (scalar type) near field we get

$$\frac{1}{r}\frac{d^2}{dr^2}(r\Phi_s) + \gamma_0 \lambda'^2 \Phi_s = -4\pi a' \Phi'_0 \delta(r), \tag{B7}$$

By comparing (B7) and (B4), obviously $\gamma_0 = p'^2$. It may seem that both equations have the same format as the Klein Gordon equation, which has erroneously been derived for fermions as a fore-runner of Dirac's Equation. The presence of the source term, however, make these two bosonic equations different.

It will be clear now that, under the modification (B3), the Stueckelberg mechanism allows to give a spatial description of the functionally defined Higgs field. The field can be assigned to a single composite source that produces a vectorial Proca type (weak force) far field and a scalar type (strong force) near field. For reasons of simplicity, I wish to stick to the two-parameter formulation as expressed by (38) rather than by the four-parameter equivalent (B4).